\documentclass[a4paper,fleqn,usenatbib]{mnras}

\usepackage{newtxtext,newtxmath}

\usepackage[T1]{fontenc}
\usepackage{ae,aecompl}

\usepackage{graphicx}	
\usepackage{amsmath}	

\title[CGM Cold Gas Clumpiness]{{Clumpiness of Observed and Simulated Cold Circumgalactic Gas}}
\author[R. Augustin et al.]{Ramona Augustin$^{1}$\thanks{E-mail: raugustin@stsci.edu},
C{\'e}line P{\'e}roux$^{2,3}$,
Aleksandra Hamanowicz$^{2}$,
Varsha Kulkarni$^{4}$,
\newauthor
Hadi Rahmani$^{5,6}$,
Anita Zanella$^{7}$
\\
$^{1}$Space Telescope Science Institute, 3700 San Martin Drive, Baltimore, MD, 21218, USA\\
$^{2}$European Southern Observatory, Karl-Schwarzschildstrasse 2, D-85748 Garching bei M{\"u}nchen, Germany\\
$^{3}$Aix Marseille Universit\'e, CNRS, LAM (Laboratoire d'Astrophysique de Marseille) UMR 7326, 13388, Marseille, France \\
$^{4}$Dept. of Physics and Astronomy, Univ. of South Carolina, Columbia, SC 29208, USA\\
$^{5}$GEPI, Observatoire de Paris, PSL Research University, CNRS, Place Jules Janssen, 92190 Meudon, France\\
$^{6}$School of Astronomy, Institute for Research in Fundamental Sciences (IPM), PO Box 19395-5531 Tehran, Iran\\
$^{7}$Istituto Nazionale di Astrofisica, Padova Astronomical Observatory, Vicolo dell’Osservatorio 5, 35122 Padova, Italy \\
}

\date{Accepted XXX. Received YYY; in original form ZZZ}

\pubyear{2020}

\begin{document}
\label{firstpage}
\pagerange{\pageref{firstpage}--\pageref{lastpage}}
\maketitle

\begin{abstract}

{Determining the clumpiness of matter around galaxies} is pivotal to a full understanding of the spatially inhomogeneous, multi-phase gas in the circumgalactic medium (CGM). We {combine high spatially resolved 3D observations with  hydrodynamical cosmological simulations to measure the cold circumgalactic gas clumpiness}. We present new adaptive-optics-assisted VLT/MUSE observations of a quadruply lensed quasar, targeting the CGM of {2} foreground $z\sim$1 galaxies observed in absorption. We {additionally} use zoom-in FOGGIE simulations with exquisite resolution ($\sim$0.1 kpc scales) in the CGM of galaxies to compute {the physical properties of cold gas traced by Mg\,II absorbers}. {By contrasting these mock-observables with the VLT/MUSE observations, we find a large} spread of fractional variations of Mg\,II equivalent widths with physical separation, both in observations and simulations. The simulations indicate a dependence of the Mg\,II coherence length on the underlying gas morphology (filaments vs clumps). {The} $z_{\rm abs}$=1.168 Mg\,II system {shows coherence} over $\gtrsim$ 6 kpc and is associated with an [O\,II] emitting  galaxy situated 89 kpc away, with SFR $\geq$ 4.6 $\pm$ {1.5} $\rm M_{\odot}$/yr and $M_{*}=10^{9.6\pm0.2} M_{\odot}$. {Based on this combined analysis, we determine that} the absorber is consistent with being an inflowing filament. The $z_{\rm abs}$=1.393 Mg\,II system traces dense CGM gas clumps varying in strength over $\lesssim$ 2 kpc physical scales. Our findings suggest {that} this absorber {is likely related to} an outflowing clump. Our joint approach combining {3D-spectroscopy} observations of lensed systems and simulations {with extreme resolution in the CGM} put new constraints on the {clumpiness of cold} CGM gas, a key diagnostic of the baryon cycle.

\end{abstract}




	\section{Introduction}

In recent years, it has become well established that the circumgalactic medium (CGM) plays a major role in galaxy evolution (see review in \citealt*{Tumlinson2017}). Pristine gas from the cosmic filaments traverses the CGM as it is accreted onto galaxies, fueling star formation. Simultaneously, on-going star formation, supernovae and active galactic nuclei drive hot, metal-enriched outflows from the galaxy which mix with and affect the pristine gas inflows. Hence the gas in the CGM is directly affected by and regulates the central galaxy's evolution. Mapping the {phase and structure} in the halos of galaxies thus provides a key diagnostic of the baryon cycle taking place in the CGM. 
	
However, due to its diffuse nature, {directly mapping those CGM structures in emission} has remained {challenging} \citep{Bertone2010a, Augustin2019, Wijers2019}. 
Recording emission line flux maps of the CGM is possible for the brightest observable emission lines (typically Ly$\alpha$ around quasars at high redshift, e.g. \citealt{Cantalupo2014,Arrigoni2019}). Only recently, deep observations at the Very Large Telescope (VLT) have allowed us to map the Ly$\alpha$ emission around normal galaxies \citep{Wisotzki2016, Leclercq2017,Leclercq2020}. The metal enrichment and {clumpiness} of these halos are however not constrained by these Ly$\alpha$ observations. 
	
In a complementary approach, absorption studies using bright background sources, such as quasars, have proven to be successful in detecting the diffuse gas in the CGM. This technique provides insight into the average composition and statistical distribution of gas around galaxies through stacking (e.g. \citealp{York2006,Bordoloi2011,York2012,Khare2012, Rudie2012}) or averaging ensembles of absorber properties (e.g. \citealp{Adelberger2005,Werk2014, Peroux2014}). These studies report a characteristic drop of metal line strength with increasing impact parameter.
Such statistical studies provide key information regarding the averaged CGM {cold gas} distribution around different galaxy populations, {where cold gas refers to 10$^4$K gas (as opposed to molecular).}
However, one of the drawbacks of absorption studies is that the information gained is limited to a {isolated} pencil beam along a single line of sight.

To further constrain the CGM structure around galaxies, alternative approaches have to be used. Studies of absorbers along close quasar pairs have provided new information on the {clumpiness} of the gas in single systems {probed by adjacent lines-of-sight} \citep{Hennawi2006, Martin2010, Rubin2015}. These works infer that the high-ionization material arises predominantly in large, quiescent structures extending beyond the scale of the absorber host dark matter halos rather than in ongoing galactic winds. On smaller scales, gravitationally lensed background objects provide multiple images as background sources to study the material in absorption (e.g. \citealp{Rauch2001, Ellison2004, Lopez2007, Chen2014, Rubin2018,Kulkarni2019}). Such observations allow us to determine the coherence scale which is the length scale over which the gas traced by Ly$\alpha$ or metals does not vary. \cite{Rubin2018} suggest that even weak Mg\,II absorbers show a high degree of coherence over large scales ($\sim$8$-$22 kpc). Taken together, these observations indicate that the coherence
length of C\,IV systems is larger than that of Mg\,II
systems. This is qualitatively consistent with the simple picture of clumpy, low ionisation,
gas embedded in larger, more homogeneous, and more highly
ionised outer halos. In the gravitational lens approach, the range
of linear scales probed is directly related to the instrument spatial resolution capabilities.
More recently, it has become possible to use extended background sources (e.g. \citealt{Lopez2018, Peroux2018, Mortensen2020}) to probe {cold gas} over continuous areas on sub-kpc scales. These results confirm the trend of absorption strength decreasing with increasing impact parameter, in agreement with the statistics towards quasars \citep{Lopez2018, Mortensen2020, Lopez2020}. Specifically, \cite{Peroux2018} find a good efficiency of {cold gas} mixing on kpc-scales in the CGM of a typical $z\sim$1 galaxy.

{From a theoretical view-point, the ability of cosmological hydrodynamical simulations to reproduce the column
 density distribution of H\,I, the line widths and profiles tracing the
 temperature of the gas, and the evolution of the line density with redshift has
 been one of the great successes of the last few decades \citep[][but see
 \citealt{Gaikwad2017}]{Rahmati2013}. Such simulations have provided insight into the broad physical environments within and between galaxies probed by various H\,I column density regimes, and thus allowed us to assess the implications of their observational properties. However, cosmological hydrodynamical simulations are still limited by their resolution, as most of the physical processes which play a key role in galaxy evolution (e.g. star formation and its feedback, black hole growth and its feedback) occur at scales well below what can be simulated. There have been numerous recent efforts to increase the resolution in the typically underresolved CGM regions of galaxies \citep{Hummels2018,vandevoort2019,Peeples2019,suresh2019}. 
 Indeed, this step is key to simulate the cold phase ($\sim 10^4$ K) of the gas traced by e.g. Mg\,II absorbers in observations. For these reasons, efforts in the last few years have mostly focused on the higher-ionisation phase of the CGM \citep{Ford2016, Gutcke2017,Suresh2017, Muratov2017,Corlies2020}. 
 The most recent high-resolution cosmological simulations now allow to study the cool gas content and clumpiness of the CGM (e.g. \citealp{Peeples2019,Zheng2020,Nelson2020,vandeVoort2020}).
}
    
In this work, we make use of both Adaptive Optics (AO) supported VLT/MUSE observations with an image quality of 0.7 arcsec FWHM and zoom-in cosmological FOGGIE \citet{Peeples2019} simulations with exquisite resolution ($\sim$ 0.1 kpc scales) in the CGM region to put new constraints on the {cold gas clumpiness} in the CGM of $z\sim$1 galaxies.
This work is structured as follows: Section 2 presents the observations used for this study. Section 3 details the analysis of the observations, and in Section 4 we {introduce the simulations}. 
{The analysis and interpretation of the nature of the cold CGM gas is discussed} in Section 5, before we conclude in Section 6.
Throughout this work we assume a flat $\Lambda$CDM cosmology with $\Omega_{m}$ = 0.3 and $H_0$ = 68 km/s/Mpc.

	\section{New AO-MUSE observations}
	
\begin{figure}
	\centering
	\includegraphics[width=.5\textwidth]{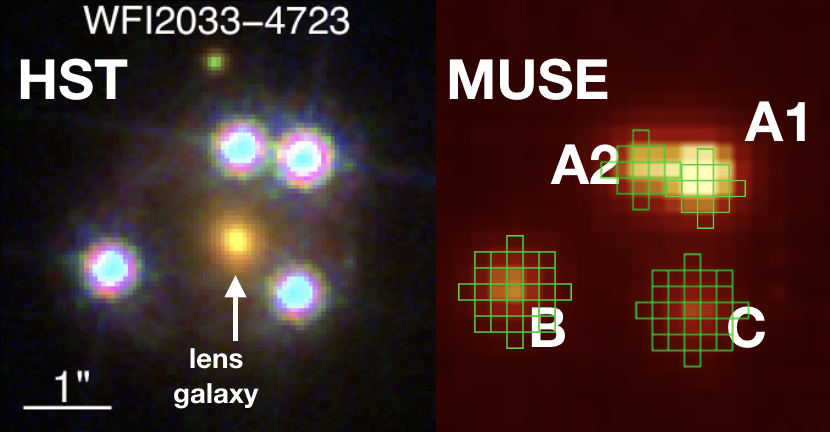}
	\caption[Zoom in to the center of Figure \ref{muse_white_light}]{{\it Left Panel:} HST broad-band composite image from three filters in the optical and near-infrared \citep{Suyu2017} of $z_{\rm QSO}$=1.66 WFI2033--4723. The lensed quasar produces four distinct images. Also apparent in this image is the $z_{\rm lens}$=0.657 lens galaxy. {Given the lensing effect, the separation between B and C is 5 kpc at $z$=1.168 and 2.3 kpc at $z$=1.393, corresponding to the intervening Mg\,II systems.} {\it Right Panel:} MUSE narrow-band observations of the same system with labels for the individual images as used throughout this work. The green contours illustrate the pixels used to extract the spectrum of each quasar image. While the spatial resolution offered by the HST is unmatched by the ground-based data, MUSE Integral Field Spectroscopy aided with the AO-system uniquely allows us to recover the spectral information. 	}
	\label{imagelabels}
	\end{figure}

The quasar WFI 2033$-$4723 ($J$ = 16.3 mag and J2000 coordinates: 203342.12--472343.9) lies at a redshift $z_{\rm QSO}$=1.66 \citep{Morgan2004,Eigenbrod2006}. The source is known to be lensed by a foreground galaxy at a spectroscopically determined redshift of $z_{\rm lens}$=0.657 \citep{Ofek2006,Eigenbrod2006,Sluse2019} resulting in a system of four distinct images. The left panel of Figure \ref{imagelabels} displays the Hubble Space Telescope (HST) broad-band imaging illustrating the configuration of the system: four distinct quasar images with the lens galaxy clearly located in the center. 
To achieve a spatial resolution which allows for separation of the four images, we took advantage of the Science Verification (SV) program of the  in the wide-field (60 arcsec $\times$ 60 arcsec) mode with MUSE. The data were taken on 17 September 2017 in the nominal mode (Program 60.A-9189, PI: R. Augustin). 
The MUSE field was observed for 1h and was rotated by 90 deg between the exposures to minimize flat-fielding imperfections.

To improve on the depth of the observations we added (in a seeing-weighted way - see description of the reduction) archival MUSE data of the same field.
The field had been observed in natural-seeing wide-field mode with MUSE in several visits between 2014 and 2016 for a total of 5.4h (Program 097.A-0454, PI: D. Sluse and Science Verification 60.A-9306) with the goal to improve on measurements of the Hubble constant $H_0$ \citep{Suyu2017,Sluse2019}. 
In order to benefit from the deepest possible data, we have fully reprocessed and combined both the archival and {our SV} data. All the data were reduced with version 2.6 of the ESO MUSE
pipeline \citep{Weilbacher2015} as well as custom-developed codes. We run the {\it muse\_scibasic} and {\it muse\_scipost} recipes for each sub-cubes. These functions perform all the necessary data reduction steps, including correction for bias, dark, bad pixels as well as {air} wavelength and geometry calibration, baryocentric correction, flux calibration, first-step telluric sky subtraction and astrometry. Further optimised sky subtraction was carried out with the Zurich Atmosphere Purge (ZAP) software \citep{Soto2016}.
{Since the MUSE spectra are calibrated at air wavelength, we use air rest wavelengths throughout our analysis to determine redshifts.}

First, we run {\sc Sextractor} \citep{Bertin1996} on the MUSE white light image to create a segmentation map. This is then used to mask all objects in the field in order to select the sky regions. The ZAP software then performs a Principle Component Analysis on the resulting sky spectra which is further subtracted from the cube. Finally, we combined all the individual data cubes {by assigning longer exposure times and the better seeing values a larger weight}, in order to achieve the optimal depth and spatial resolution in the final data product. The resulting cube has a spatial sampling of 0.2 arcsec per pixel, a spectral sampling of 1.25 \AA\ per pixel and a wavelength coverage extending from 4750 to 9350 \AA . The image quality of the final combined data was measured from the quasar images. The resulting point spread function has a full width at half maximum (FWHM) of 0.7 arcsec at 7000\AA, as measured on the central quasar (image B). The right panel of Figure \ref{imagelabels} shows the MUSE narrow-band image of the lensed system, indicating the labels for the four images (labelled A1, A2, B and C) and showing in green contours the pixels that were extracted for each image's spectrum.

	\section{Analysis of Integral Field Spectroscopic Data}

\subsection{Quasar Spectroscopy and Absorption Features}	

\begin{table}
\caption{Measurements for the two Mg\,II absorption systems (at $z_{\rm abs}$=1.168 and $z_{\rm abs}$=1.393): {SNR}, Equivalent Widths (EW) in units of \AA\ and in the case of the $z_{\rm abs}$=1.168 system, the velocity offsets (in km/s) between the absorber sightlines and the [O\,II] emitter. }
\centering
\begin{tabular}{ccccccc}
\hline
QSO images 	&	A1 	&	A2 	&	B  	&	C  	\\	\hline	\hline
\multicolumn{5}{c}{{$\mathbf{z_{\rm abs}}$=1.393}}\\		\hline
\multicolumn{5}{c}{SNR at the absorber}									\\	\hline	
 SNR	&	 129          	&	 81          	&	 106 	&	 60	\\		
\hline
\multicolumn{5}{c}{Rest Equivalent Widths}									\\	\hline	
 EW$_{2796}$ [\AA]	&	 < 0.1          	&	 < 0.1          	&	 0.05 $\pm$ 0.01 	&	 0.41 $\pm$ 0.04 	\\		
 EW$_{2803}$ [\AA]	&	 < 0.1         	&	 < 0.1          	&	 0.03 $\pm$ 0.01  	&	 0.32 $\pm$ 0.03  	\\	\hline	
\multicolumn{5}{c}{{$\mathbf{z_{\rm abs}}$=1.168} }	\\	
\hline
\multicolumn{5}{c}{SNR at the absorber}									\\	\hline	
 SNR	&	 40          	&	 35          	&	 34 	&	 32	\\		
\hline
\multicolumn{5}{c}{Rest Equivalent Widths}									\\	\hline	
 EW$_{2796}$ [\AA]	&	 0.09 $\pm$ 0.01 	&	 0.08 $\pm$ 0.01 	&	 0.15 $\pm$ 0.02 	&	 0.05 $\pm$ 0.01 	\\		
 EW$_{2803}$ [\AA]	&	 0.05 $\pm$ 0.01 	&	 0.03 $\pm$ 0.01 	&	 0.07 $\pm$ 0.02 	&	 0.02 $\pm$ 0.02 	\\	\hline	
\multicolumn{5}{c}{Velocity Offset to [O\,II] emitter at z=  1.16895 $\pm$ 0.00007}  \\		\hline
$\Delta v$ [km/s]	&	 20 $\pm$ 50	&	 70 $\pm$ 50	&	 70 $\pm$ 50	&	 110 $\pm$ 50	\\	\hline
\end{tabular}
\label{tab:onetable}
\end{table}

\begin{figure}
\centering
\includegraphics[width=\linewidth]{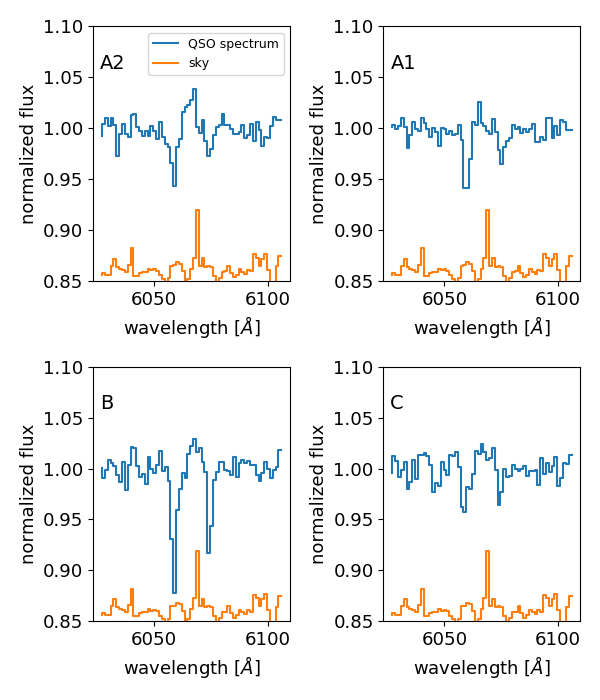}
\caption[z=1.168 absorber]{Mg\,II absorption features against the four quasar images in the $z_{\rm abs}$=1.168 absorption system. Both lines of the Mg\,II doublet are clearly detected in all four spectra. {The offset orange lines show the sky spectrum.} {The spike between the doublet is a noisy residual from the sky subtraction which is masked in the subsequent analysis.
}}
\label{fig:EW1167}
\end{figure}

\begin{figure}
\centering
\includegraphics[width=\linewidth]{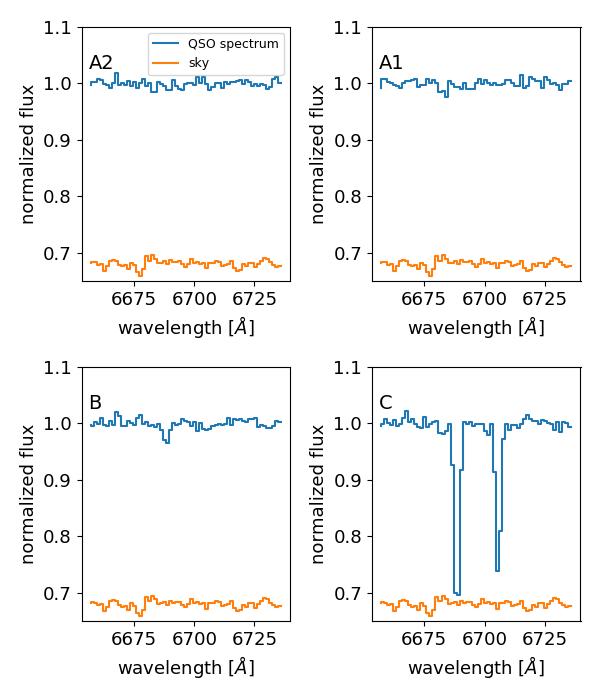} 
\caption[z=1.393 absorber]{Mg\,II absorption features against the four quasar images in the $z_{\rm abs}$=1.393 absorption system. The Mg\,II absorption is detected against quasar images C and B, and remains undetected in the two remaining spectra (A1 and A2). {The offset orange lines show the sky spectrum.}}
\label{fig:EW1392}
\end{figure}

		We obtain the spectrum of the four individual quasar images from the MUSE data cube.  
		We extract the flux from carefully selected pixels (see green mask in Figure \ref{imagelabels}), constructing an area which encompasses {at least} the FWHM of the point spread function (PSF).
		{For the two fainter images (B and C), an aperture size of 7 px = 1.4 arcsec in diameter maximises the signal-to-noise ratio (SNR) of the resulting spectra. For the  brighter counter-images A1 and A2, a smaller aperture of 5 px = 1 arcsec in diameter was used to avoid overlap between the two nearby images. 
		 The apertures are centered {by eye} on the brightest pixels for each quasar image. A {small, yet inconsequential} offset {of 1 px} from the brightest pixel for image A1 is applied, to minimise the overlap with image A2. 
		 {We want to note that for our absorption measurements the chosen area is not critical.}
		 {The chosen apertures as seen in QFitsView\footnote{https://www.mpe.mpg.de/{$\sim$}ott/dpuser/qfitsview.html} are shown in Figure  \ref{imagelabels}.}
		 }
	While we take care in the extraction of the spectra to have separate spectra for A1 and A2, we do note that due to their proximity, the spectra are likely not completely independent.

Given the redshift of the lens galaxy ($z_{\rm lens}$=0.66), the MUSE spectra cover absorption features of Mg\,I, and Ca H+K lines. We report no detections for any of these absorption lines in any of the four extracted spectra. 
We further undertake a blind search for additional absorbers and detect two Mg\,II systems: one at $z_{\rm abs}$=1.168 and one at 1.393 (Figures~\ref{fig:EW1167} and \ref{fig:EW1392}). The lower redshift absorber is detected in the four quasar spectra. The latter system is detected solely against quasar images C and B and remains undetected towards the two remaining images (A1 and A2). Figures~\ref{fig:EW1167} and \ref{fig:EW1392} show these Mg\,II absorption systems. 
We {calculate the rest-frame equivalent width of each of these systems {by integrating the line profiles} following the standard definition: }
\begin{equation} \label{eq:EW}
\rm     EW(F, \delta F) = (1+z)^{-1} \, \int (1 - ((F \pm \delta F)/F_{cont}) \, d\lambda
\end{equation}

{where $\rm F/F_{cont}$ is the continuum normalized flux of the spectrum.} For detected absorption features, we measure the EW for each Mg\,II line (2796, 2803\AA) in the normalised spectra. In order to remove the signatures intrinsic to the quasar, a local power-law continuum is fitted. Dividing the observed spectra with their associated continua results in four normalised quasar spectra corresponding to each of the four images. 
{The errors on the EWs are calculated from flux uncertainties and propagating the errors in equation \ref{eq:EW}. }

For the non-detections, we calculate the corresponding 5$\sigma$ {($\sigma$ in eq. \ref{eq:limits})} limit of EW$_{\rm Mg\,II}$ (see Table~\ref{tab:onetable}) using the relation between the observed-frame equivalent width ($\rm EW_{limit}$), the resolving power (R), and the SNR per resolution element \citep{Menard2003} {at a given observed wavelength ($\rm \lambda_{obs}$)}:

\begin{equation} \label{eq:limits}
\rm    EW_{limit} [\text{\AA}]< \sigma* \dfrac{\lambda_{obs} [\text{\AA}]}{R \times SNR}
\end{equation}

Table~\ref{tab:onetable} lists the corresponding {spectral SNR}, rest-frame EW of Mg\,II from detections and limits from non-detections for each of the absorption systems.

\subsection{Emission Galaxy Spectroscopy}

\begin{figure}
\centering
\includegraphics[width=0.7\linewidth]{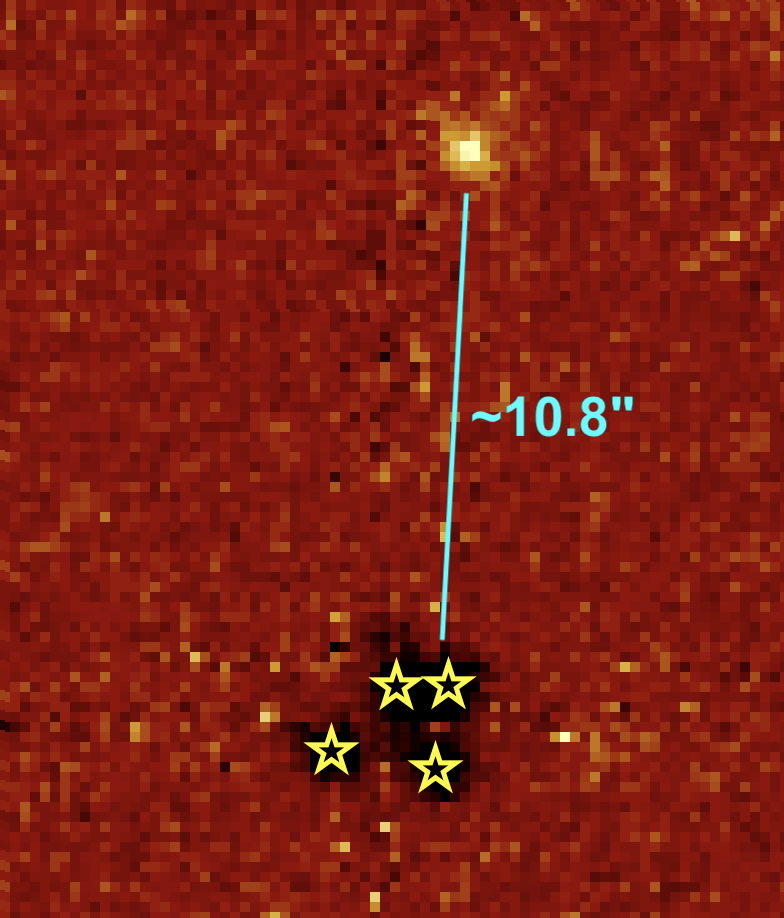}
\caption[OII emitter]{MUSE narrow-band image around the [O\,II] emission line at $z_{\rm abs}$=1.168. The lensed quasar images are subtracted, showing negative flux values at the position indicated by the yellow stars. The [O\,II] emission line is shown in light colour at an impact parameter of 10.8 arcsec (89 kpc) North of the lensed system.}
\label{10arcsec}
\end{figure}

\begin{figure}
\centering
\includegraphics[width=0.85\linewidth]{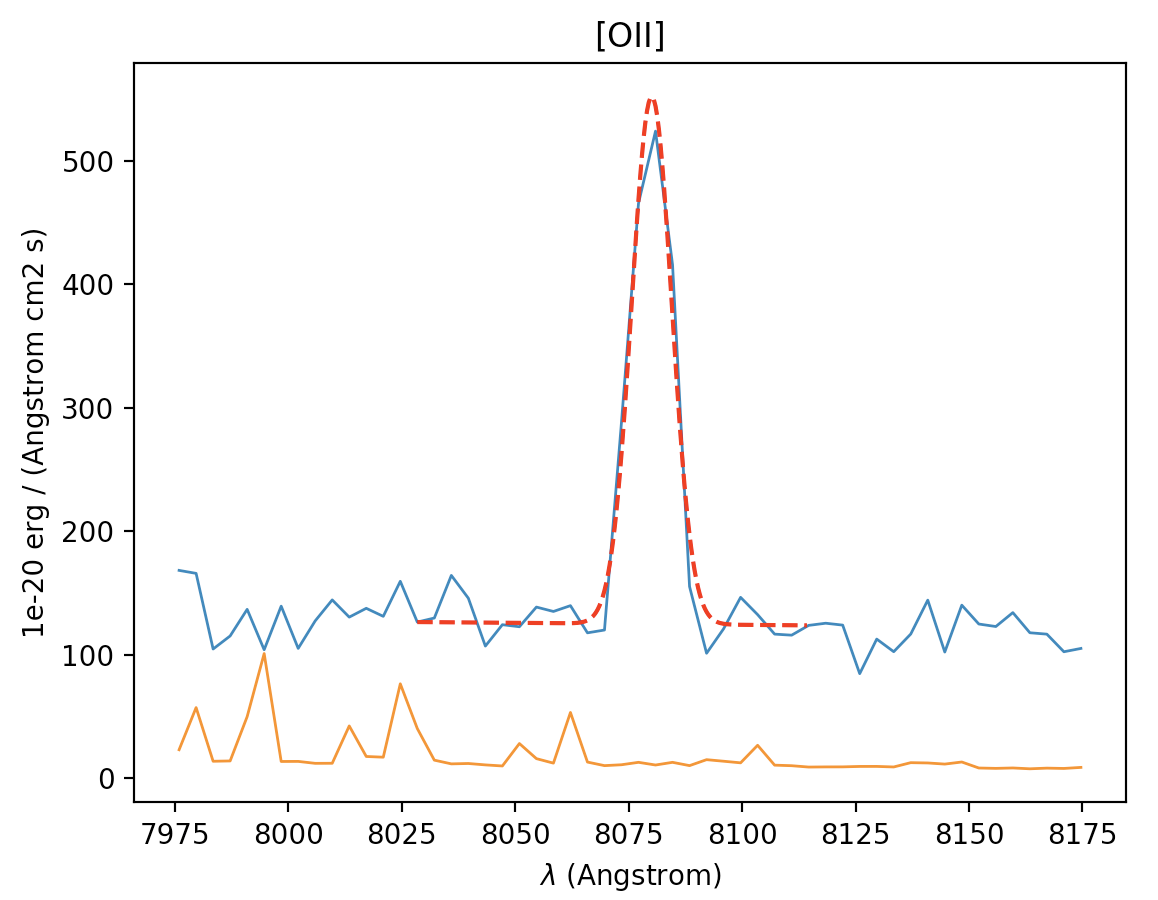}
\caption[OII fit]{MUSE spectrum of the galaxy associated with the $z_{\rm abs}$=1.168 Mg\,II absorber. The blue line shows the extracted [O\,II] emission line, binned with a 3-pixel kernel. At MUSE spectral resolution, the doublet structure of [O\,II] lines is not well-separated, which results in smearing of the different velocity components. The red line indicates the Gaussian fit to the emission line. The orange line displays the sky spectrum in the MUSE cube, also binned with a 3-pixel kernel and arbitrarily scaled in the y-direction.}
\label{fig:OIIfit}
\end{figure}

\begin{figure*}
\centering
\includegraphics[width=16cm, height=7.5cm]{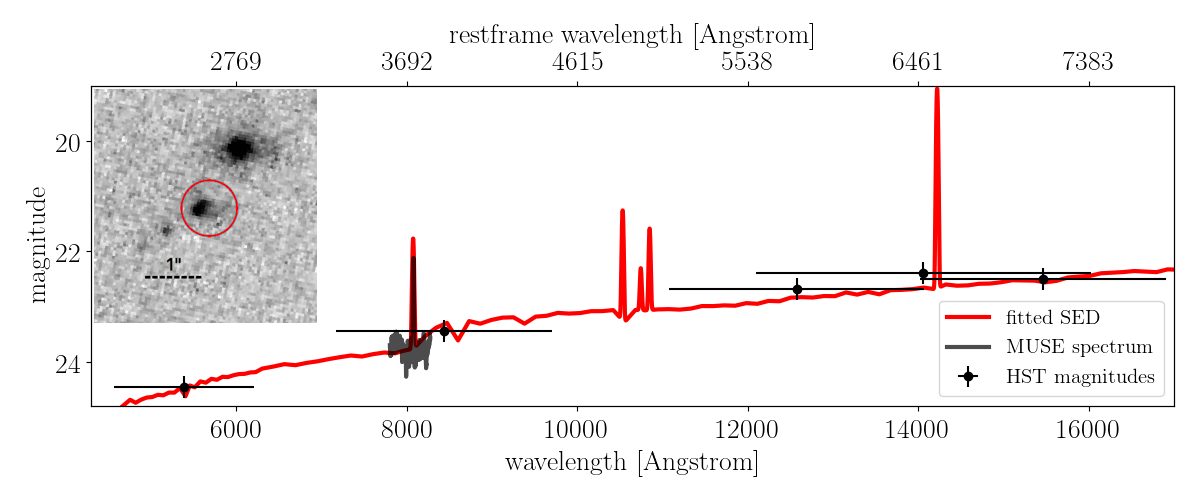}
\caption[SED fit]{Spectral Energy Distribution (SED) fit to the galaxy associated with the $z_{\rm abs}$=1.168 Mg\,II absorption system. The points with error bars indicate the continuum magnitudes observed with HST at optical and near infrared wavelengths (F555W, F814W, F125W, F140W and F160W). The red spectrum is the result from the SED fit with LePhare code \citep{Arnouts1999,Ilbert2006} leading to a stellar mass of the counterpart galaxy of $M_{*}=10^{9.6\pm0.2} M_{\odot}$. We overplot the MUSE spectrum in grey which precisely matches the spectrum resulting from the SED fit. The inset on the top-left corner shows the HST continuum V-band image of the galaxy. {The galaxy within  the red circle is detected at the absorber redshift, while the other objects are below the detection limit or at a different redshift.}}
\label{sedfit}
\end{figure*}

Integral field spectroscopy (IFS) is exceptionally well-suited for studies of multiply lensed images as well as the associated foreground galaxies. The high-spatial resolution achieved thanks to the adaptive optics coupled with the spectral resolution of MUSE means that independent spectra can be extracted for each of the images of the lensed quasar (see section above). Furthermore, the 3D information allows us to search in emission for galaxies associated with both the lens and the identified Mg\,II absorption systems. 
{The search for emission line counterparts is performed by eye in a narrow-band image centered on the expected emission line.}

At the redshift of the lens (z$_{\rm lens}$=0.657), [O\,II] and [O\,III] emission lines are covered by the MUSE observations. We do not detect these emission lines in the MUSE cube {down to a SFR threshold of 0.01 $\rm M_{\odot}$/yr}. 
At the redshift of the $z_{\rm abs}$=1.393 Mg\,II system, the [O\,II] emission line is also covered by the MUSE observations, {but the [O\,II] 3729 line lies close to a sky emission line which results in additional contamination. We do not report  counterpart galaxies to this absorbers down to a SFR detection limit of 0.02 $\rm M_{\odot}$/yr.}
We find a counterpart galaxy at the redshift of the $z_{\rm abs}$=1.168 Mg\,II system. The [O\,II] emission line is detected to the North of the quasar images at a distance of 10.8 arcsec, corresponding to 89 kpc at the redshift of the absorber (see Figure~\ref{10arcsec}). At the MUSE spectral resolution, the doublet structure of [O\,II] lines is not well-separated when extracting the spectrum integrating over the whole galaxy, which results in smearing of the different velocity components (see also \citealt{Hamanowicz2020}). To {calculate} the total flux in the emission line, we fit a gaussian function to the spatially integrated [O\,II] profile (shown in Figure \ref{fig:OIIfit}). From the [O\,II] emission flux, we determine the star formation rate (SFR) of the galaxy to be SFR = 4.6 $\pm$ {1.5} $\rm M_{\odot}$/yr following \cite{Kennicutt1998}.
This SFR is not dust corrected, as we have no accurate means of determining the dust correction, and therefore presents a lower limit on the true SFR of this galaxy.
{The SFR detection limit in the data {calculated from the sky level} for that redshift is 0.01 $\rm M_{\odot}$/yr.}

The galaxy is also detected in continuum in archival HST images. The MAST data archive provides the magnitudes of the galaxy in 5 different bands at optical and near infrared wavelengths (F555W, F814W, F125W, F140W and F160W). We use the magnitude determination in these bands to perform a Spectral Energy Density (SED) fit of the stellar population in this object. To this end, we utilize the SED fitting code LePhare \citep{Arnouts1999,Ilbert2006}. The results of the SED fit, shown in Figure \ref{sedfit}, matches precisely the MUSE galaxy spectrum where they overlap. The fit indicates a stellar mass of the counterpart galaxy of $M_{*}=10^{9.6\pm0.2} M_{\odot}$.
{We note that there are other objects closeby in projection, but they are either at a different redshift or are not detected in MUSE data cube. }
We further run {\sc Sextractor} \citep{Bertin1996} on the HST V-band (F555W) image in order to determine the position angle of the galaxy to be PA=$-$48$\pm$41 degrees. 
The error is large due to the round appearance of the galaxy. 
We derive the "azimuthal angle" between the image A1 quasar line-of-sight and the projected galaxy’s major axis on the sky to be 45$\pm$41 degrees. 
We can therefore not constrain the absorber alignment with respect to the galaxy orientation.
We use the redshifts resulting from {the centroids of} the [O\,II] emission line as well as the Mg\,II absorption lines at $z_{\rm abs}$=1.168 to calculate the velocity offset of the various absorption features to the systemic redshift of the host galaxy. 
{The centroids were determined from a first moment analysis using astropy specutils centroid function and verified by eye.}
{The errors on the velocities include wavelength calibration uncertainties of $\sim$ 20 km/s.} The results are tabulated in Table~\ref{tab:onetable}.

\subsection{Physical Distances at the Absorbers' Redshift}

To relate the physical properties of gas detected in absorption along the lines-of-sight of the quasar images to the foreground objects, we calculate the separation between the images in the absorber plane in physical units.	We perform this calculation in comoving space, denoting all following distances as comoving distances. The approach is analogous to equation 2 in \cite{Cooke2010} and equation 1 in \cite{Kulkarni2019}, who express their derivation in angular diameter distances.

At the quasar's redshift (z$_{\rm QSO}$=1.66), the separation of all the four images is naturally zero. Assuming our cosmology, we calculate the distances between two quasar images in the absorber plane ($d_{\rm abs}$) from the distances in the lens plane ($d_{\rm lens}$) in the following way:
	\begin{equation}
	d_{\rm abs} = \frac{d_{\rm abs - QSO}}{d_{\rm lens - QSO}} * d_{\rm lens}
	\end{equation}
where $d_{\rm abs - QSO}$ is the distance between the foreground absorber and background quasar and $d_{\rm lens - QSO}$ is the distance between the lens galaxy and the background quasar. The comoving distance between the lens (z$_{\rm lens}$=0.657) and the quasar (z$_{\rm QSO}$=1.66) is 1579 $h^{-1}$ Mpc. For the low-redshift Mg\,II system, the distance between the absorber ($z_{\rm abs}$=1.168) and the quasar is 669 $h^{-1}$ Mpc; while the distance between the high-resdshift Mg\,II absorber ($z_{\rm abs}$=1.393) and the quasar is 342 $h^{-1}$ Mpc. The resulting physical distances for each of the two Mg\,II systems range from 0.6 to $\sim$5 kpc and are tabulated in Table \ref{tab:EWfracs}.

\section{CGM Zoom-in Hydrodynamical Cosmological Simulations}
\label{sec:simu}
\begin{figure}
    \centering
    \includegraphics[width=\linewidth]{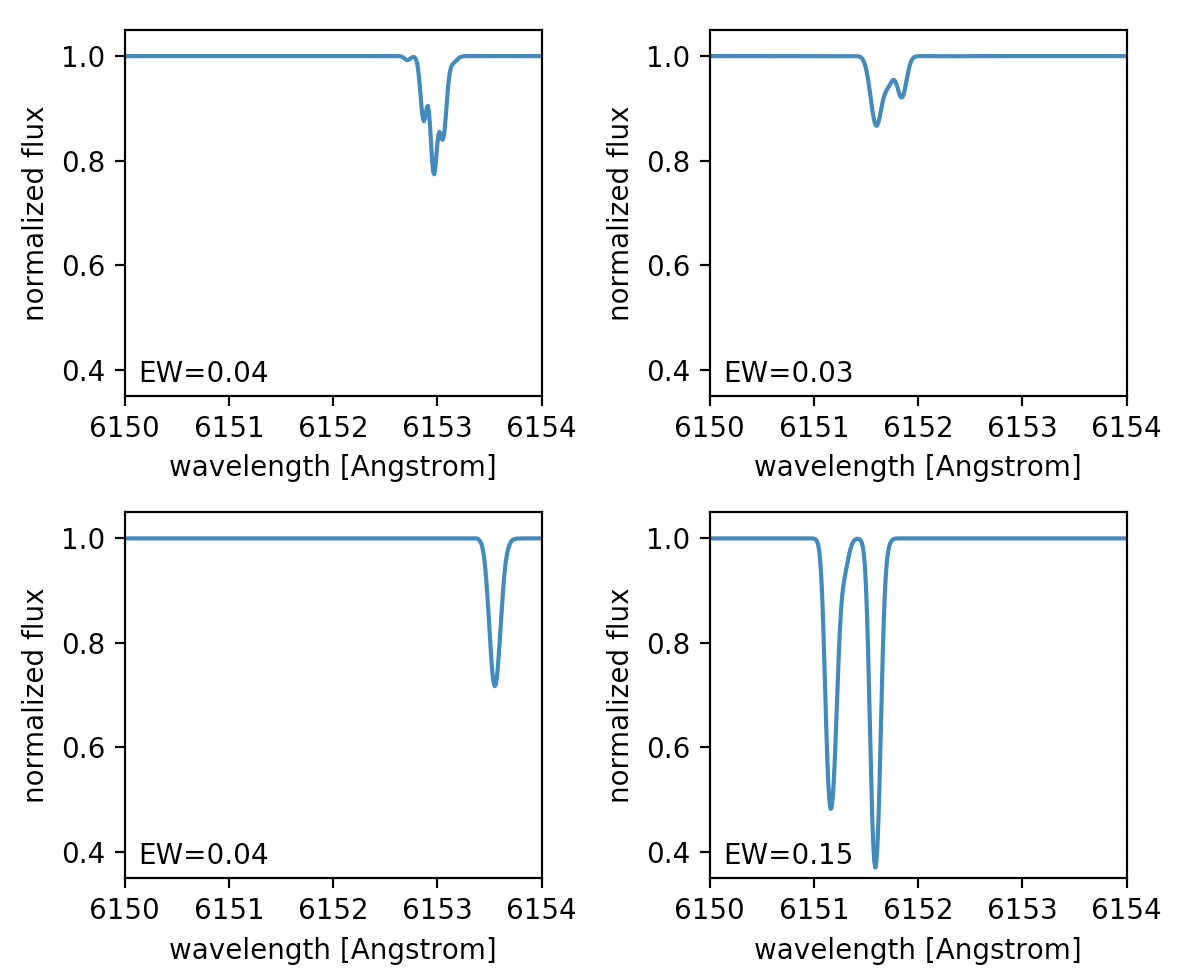}
    \caption{{Example synthetic absorption line spectrum. This mock CGM Mg\,II $\lambda$2796\AA\ systems were created using the \textsc{Trident} package to pierce random sightlines through the modelled halo from the FOGGIE hydrodynamical cosmological simulations.
    We integrate over the multiple components of an absorber to calculate the EW as those would not be resolved in our MUSE data.
    }}
    \label{fig:synthspec}
\end{figure}

\begin{figure*}
\centering
\includegraphics[width=0.7\linewidth]{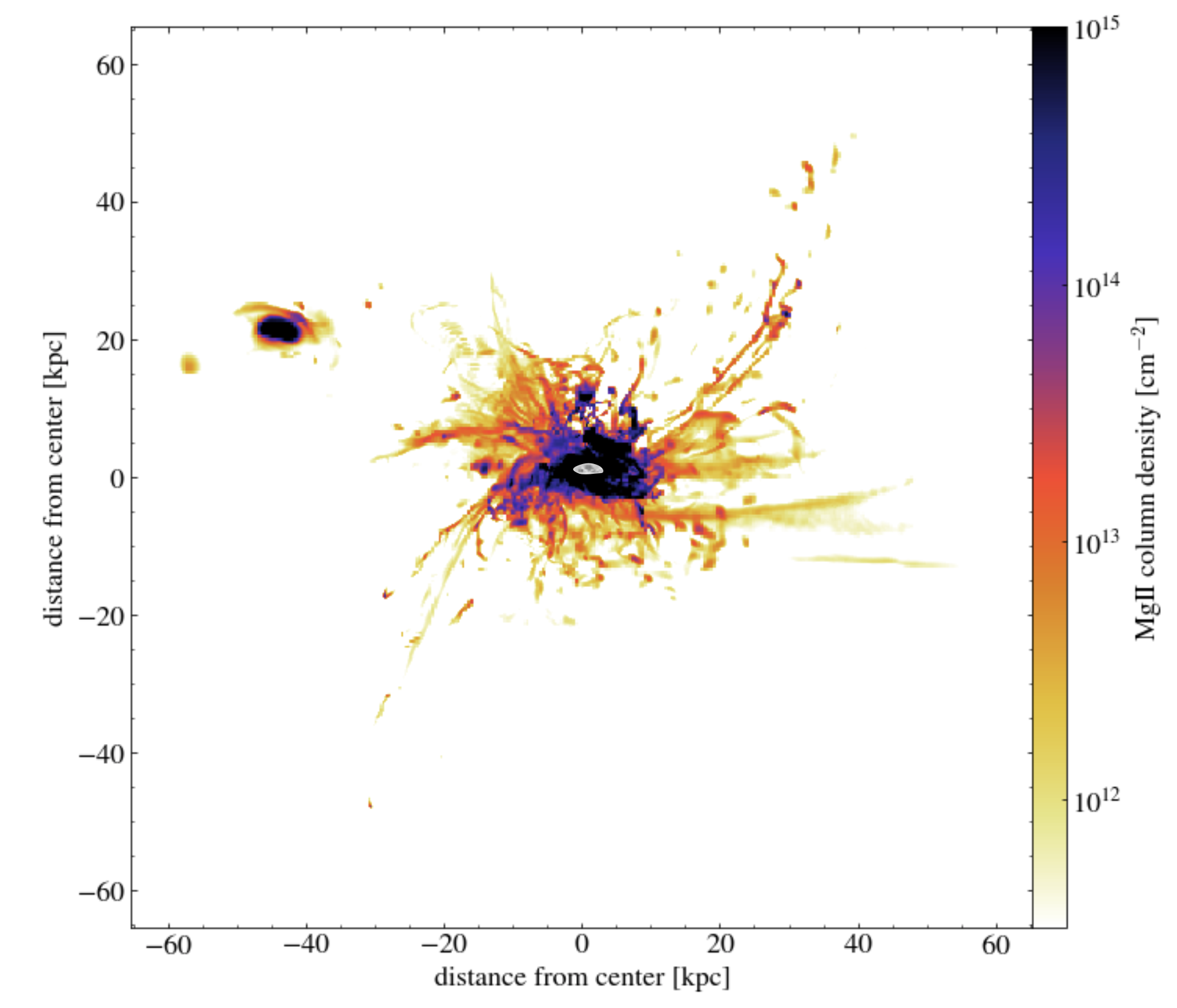}
\includegraphics[width=0.7\linewidth]{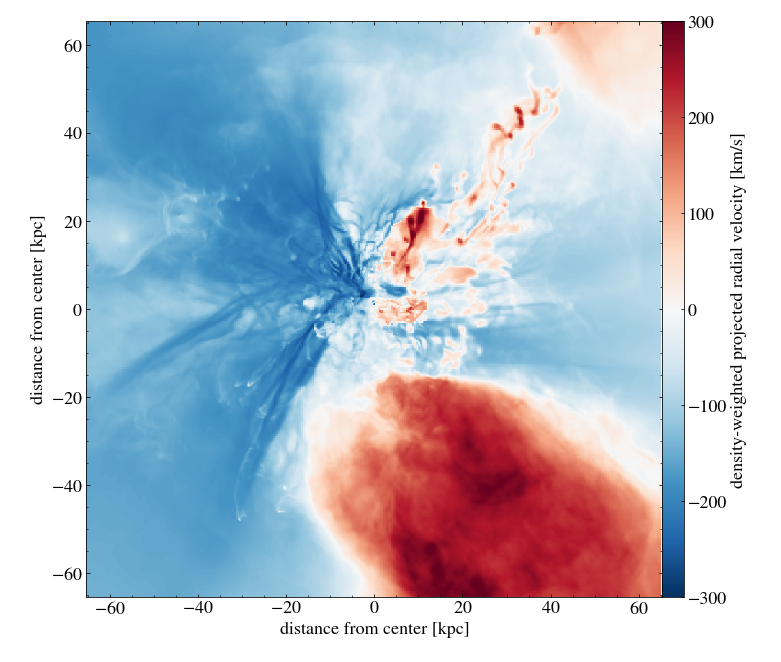}
\caption[foggie halo]{
Mg\,II halo at z=1.2 from the FOGGIE simulations. The Mg\,II column densities corresponding to the observed range are shown on the top, while the density-weighted radial velocity of the gas (inflow blue, outflow red) is displayed at the bottom.
The grey structure in the center of the upper panel depicts the density projection of the galactic disc, showing the galaxy's orientation to be close to edge-on.
This halo has a virial radius of 84 kpc, a dark matter halo mass of $\rm 10^{11.4} M_{\odot}$, a stellar mass of  $\rm 10^{10.5} M_{\odot}$, a SFR of $\rm 10 M_{\odot}/yr$, and an ISM mass of $\rm 10^{9.4} M_{\odot}$. This mock halo indicates the Mg\,II structures range from small, dense clumps, to extended, large filaments.
It also appears that extended structures with log(N$\rm _{Mg\,II}) \sim $ 12-13 are preferentially tracing inflow filaments while most clumpy structures with log(N$\rm _{Mg\,II}) > $ 13 trace dense, outflowing material {as illustrated by the radial velocity map which presents both the velocity and direction of these gas flows.} {We therefore conclude that there may be structural differences in Mg\,II clouds depending on whether they trace inflows or outflows.}
}
\label{fig:foggiehalo}
\end{figure*}

{In order to constrain the cold gas in CGM and the nature of the detected Mg\,II absorbers, 
we make use of CGM simulations with high spatial resolution. }{We use} the {\it FOGGIE} simulation suite which is designed for exquisite resolution ($\sim$ 0.1 kpc scales) in the CGM of galaxies which evolve into Milky-Way-like systems at z=0 \citep{Peeples2019}. We create mock observations from these cosmological zoom-in simulations.
For the comparison, we choose the `Tempest' halo with forced refinement in the CGM  at redshift z=1.2, to match our observations \citep[see][for further information on this simulation]{Zheng2020}.
This halo has a virial radius of 84 kpc, a Dark Matter mass of $\rm 10^{11.4} M_{\odot}$, a stellar mass of  $\rm 10^{10.5} M_{\odot}$, a SFR of $\rm 10 M_{\odot}/yr$, and an interstellar medium (ISM) mass of $\rm 10^{9.4} M_{\odot}$. The stellar mass of the simulated halo $\sim$10 times larger than the galaxies observed with MUSE. 
However, the stellar-mass halo-mass relation for isolated field galaxies (e.g. \citealt{Read2017}) suggests that our galaxy has a comparable halo mass ($\rm \sim 10^{11} M_{\odot}$) to the one in this simulation. 

{Next, we use the {\it Trident} package \citep*{Hummels2017} to create mock spectra from idealised background source piercing through the simulated halo. {\it Trident} uses the information of the gas cells from the simulation (density, temperature, metallicity, velocity) so that spatial resolution is kept the highest possible. The \cite{Morton2003} atomic data are used to determine the Mg\,II densities in each gas cell as well as the effective redshift along a given sightline.} {{\it Trident} furthermore creates Voigt profiles and returns a synthetic normalised absorption spectrum. We shoot random sightlines through the halo (excluding the ISM region of the central and satellite galaxy) to create synthetic spectra. Figure \ref{fig:synthspec} shows example spectra of CGM Mg\,II $\lambda$2796\AA\ absorbers. 
While {\it Trident} also produces noisy mock spectra, we choose to perform the analysis on noise-less outputs to recover the simulated Mg\,II properties.
{By performing our analysis on those outputs, we recover the "true" underlying distribution of EW fractional differences in the simulated galactic halo.}

{We measure the EW of such absorbers by integrating over the absorption signal, in an analogous way to the analysis performed on the observed spectra {(see eq. \ref{eq:EW})}.}
The thus measured EWs are used for comparison with the observations (see Section \ref{section:EWfrac} and Figure \ref{fig:EWfrac}).}

A Mg\,II column density projected map of this halo is shown in the upper panel of Figure \ref{fig:foggiehalo}. The range of plotted column densities is chosen such that the lower limit of the color bar is matched to the detection limit in the observations. The column densities measured in the $z_{\rm abs}$=1.168 Mg\,II system are represented by the yellow/orange part of the colorbar, whereas the higher redshift absorber lies into the dark red region. The simulations indicate the prevalence of a variety of structures: while a large fraction of the cells in the halo are below the detection limit, the structures which are above the {observations} detection limit ($\rm 10^{11.5} cm^{-2}$) range from small, dense clumps, to extended, large filaments. 
We note that the map shown in this figure does not encompass the whole region within the virial radius, but instead targets the high-resolution CGM region of the simulation. 
We focus here on the highly resolved structures which are a least 20 kpc away from the galactic center,
{as the lensed quasar images pierce random sightlines through the CGM gas of a foreground galaxy.}
The bottom panel of Figure \ref{fig:foggiehalo} shows the density weighted radial velocity projection of the halo. 
The clump like structures seen in the Mg\,II column density map are coinciding with outflowing dense gas structures, whereas the filamentary structures are tracing inflows onto the galaxy.
Towards the lower right corner there is a fast outflow which is too hot to host Mg\,II absorbers.

\section{Nature of the Cold Circumgalactic Gas}

	\subsection{{Clumpiness} traced by the Fractional Difference in Equivalent Width }\label{section:EWfrac}

\subsubsection{Results from Observations}

The Mg\,II systems observed with MUSE against four quasar images put new constraints on the {clumpiness} of cold metals in the CGM of $z\sim$1 galaxies. In order to quantify the underlying spatial structure of the Mg\,II absorbers, we calculate the fractional variation in EW to determine the coherence lengths of absorbers. {While other statistics could have been used,}  this diagnostic is commonly used in literature \citep{Ellison2004,Rubin2018} and it enables comparison with previous works. We follow the definition for the fractional difference in EW of two sightlines (X and Y), where EW(sightlineX) $>$ EW(sightlineY):

\begin{equation}
\rm
EWfrac = \dfrac{EW(sightline X) - EW(sightline Y)}{EW(sightline X)}
\label{eq:EWfrac}
\end{equation}

\begin{figure*}
\centering
\includegraphics[width=.36\linewidth]{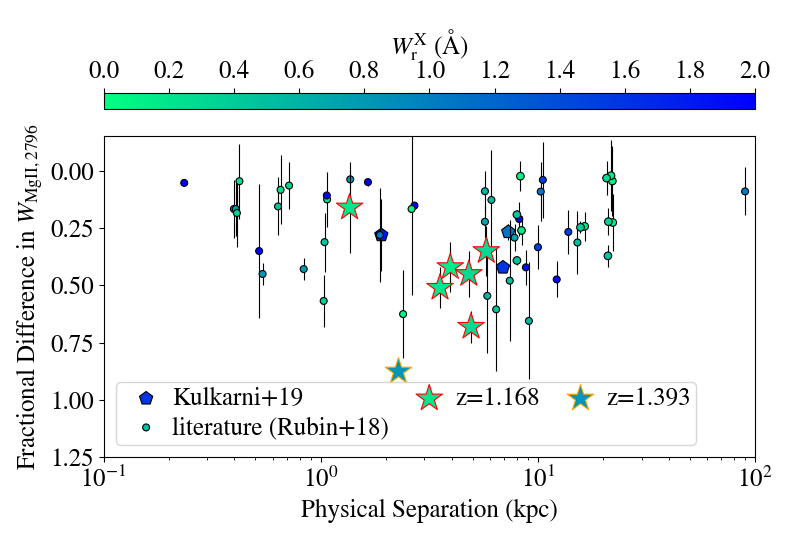}
\includegraphics[width=.36\linewidth]{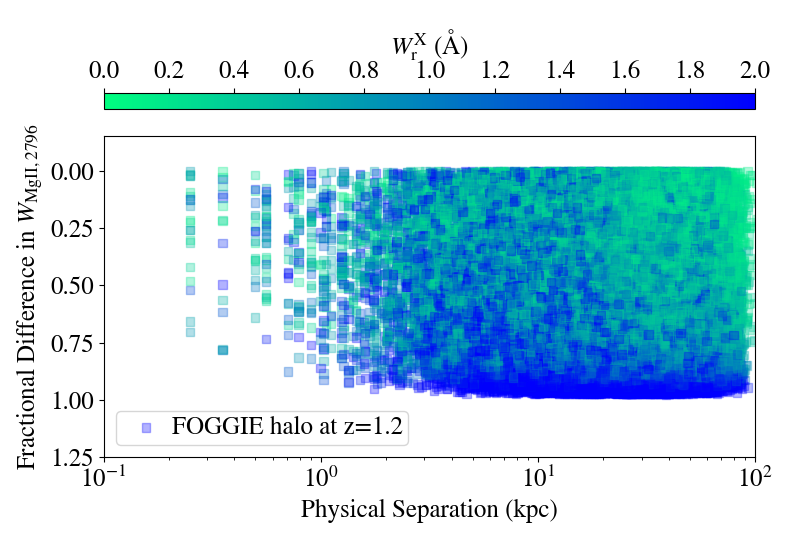}
\includegraphics[width=.27\linewidth]{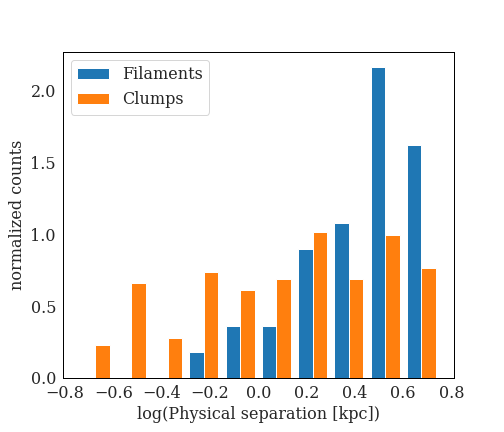}

\caption[EW diff vs separation]{Mg\,II $\lambda$2796\AA\ Equivalent Width (EW) fractional difference (see equation \ref{eq:EWfrac}) as a function of physical separation between two {random sightlines through a halo}. {\it Left panel:} Observed fractional difference from our work (stars), detections and literature sample from \cite{Rubin2018}, to which we added newer measurements \citep{Kulkarni2019}. 
The colour coding indicates the EW strength as indicated on the upper colour bar. {\it Middle panel:} Simulated fractional difference from random sightlines piercing the simulated Mg\,II halo shown in Figure~\ref{fig:foggiehalo}. We report a clear trend of high EW Mg\,II systems (blue, mainly located in the central part of the halo) showing large variation over all scales ($\sim$1-100 kpc). 
Low EW Mg\,II systems (green) display a large scatter in fractional variation at all scales, indicating that those probe a variety of structures, from clouds to filaments with large coherence lengths over tens of kpc. 
{This large coherence length is due to both the low EW systems being more extended around the central galaxy and forming larger coherent structures in the CGM.}
{\it Right panel:} Separating the halo shown in Figure~\ref{fig:foggiehalo} into structures of clumps and filaments, we show the normalized distribution of strongly (EW fractional difference > 0.5) varying systems from the middle panel for the separations up to $\sim$6 kpc. Clumps vary equally over all those scales, while filaments show a trend with separation and vary more at scales of $\gtrsim$ 3 kpc. {For this reason we conclude that the $z_{\rm abs}$=1.168 system is consistent with the signatures from a filament, while the $z_{\rm abs}$=1.393 system is consistent with the signatures from a clump.}}
\label{fig:EWfrac}
\end{figure*}

The results are tabulated in Table \ref{tab:EWfracs} and illustrated in the right panel of Figure \ref{fig:EWfrac}.
The Figure shows our data as stars, where filled symbols indicate that Mg\,II was detected towards both sightlines. 
{Due to possible cross-contamination of the spectra between sightlines A1 and A2, that fractional difference in EW is presumably biased towards lower values. {We additionally extract a spectrum between images A1 and A2 to include this component {which is of order of 0.005 \AA\ } to the error budget}.}
We also report similar measurements from the literature: detections and literature sample from \cite{Rubin2018}, to which we added the more recent measurements \citep{Kulkarni2019}. 

 \begin{table}
\caption{{Mg\,II line variation} as traced by EW fractional difference for the two Mg\,II systems in the MUSE observations.}
\centering
\begin{tabular}{ccc}
\hline
QSO Images & EW Fractional Difference & Physical Separation [kpc]\\
\hline
\multicolumn{3}{c}{{$\mathbf{z_{\rm abs}}$=1.168}}\\ \hline
A1 - A2 & 0.16$\pm$0.20 & 1.7 $\pm$ 0.2\\
A1 - C & 0.51$\pm$0.09 & 3.6 $\pm$ 0.2 \\
A2 - C & 0.42$\pm$0.11 & 4.1 $\pm$ 0.2 \\
B - A1 & 0.35$\pm$0.11  & 5.9 $\pm$ 0.2 \\
B - A2 & 0.45$\pm$0.10 & 4.8 $\pm$ 0.2 \\
B - C & 0.68$\pm$0.07 & 5.0 $\pm$ 0.2 \\
\hline
\multicolumn{3}{c}{{$\mathbf{z_{\rm abs}}$=1.393}}\\ \hline
A1 - A2 & --- & 0.8 $\pm$ 0.1 \\
B - A1 & > 0.28 & 2.7 $\pm$ 0.1 \\
B - A2 & >0.38 & 2.2 $\pm$ 0.1 \\
C - A1 & >0.91 & 1.7 $\pm$ 0.1 \\
C - A2 & >0.92 & 1.9 $\pm$ 0.1 \\
C - B & 0.87$\pm$0.02 & 2.3 $\pm$ 0.1 \\
\hline
\end{tabular}
\label{tab:EWfracs}
\end{table}

   \cite{Rubin2018} suggest that even weak Mg\,II absorbers show a high degree of coherence over large scales ($\sim$8$-$22 kpc).
The absorber at $z_{\rm abs}$=1.168 reported here is in line with this conclusion.
The Mg\,II line strength in the absorption system at $z_{\rm abs}$=1.393, however, varies strongly on kpc scales, indicating a highly inhomogeneous and clumpy structure of dense Mg\,II clouds.
This raises the question of the dependence of the coherence scale on the strength of the absorber and the geometry of the underlying CGM structure. 
While filaments will produce large coherence lengths thanks to their elongated shape, the variation in EW can be strong in the perpendicular direction.
Small dense gas clumps, individually show small coherence lengths of typically kpc scales. However depending on the sky covering fraction of small gas clumps, it may be possible, but less likely, to find two sightlines in a halo which probe individual structures of similar EWs but at large physical separation, also resulting in apparent large coherence lengths. The natural expectation thus would have Figure~\ref{fig:EWfrac} indicating a typical physical scale above which the scatter in EW fractional difference becomes important. The latest results indicate a large scatter of the EW fractional variations on scales $>$ 2 kpc.
For scales $<$ 2 kpc, fewer data points are available as those scales are close or below the typical spatial resolution scales one can probe with multiple sightlines. However, they show less scatter and typically little variation on those spatial scales.

In the quadruply lensed system reported here, the benefit of studying four closeby ($<$ 6 kpc) sightlines is that these observations are less prone to geometrical effects. 
The quasar sightlines probe absorption towards different directions (which is not possible with only a pair of sightlines) and enable us to determine the elongation and extent of the underlying structures, obtaining constraints on whether the absorber probes filamentary or clumpy structures. Indeed, the simulations introduced in the previous section demonstrate that there is a variety of structures in the CGM of $z\sim$1 galaxies, which challenges the interpretation of the fractional difference in EW depending on the orientation of probing sightlines.

\subsubsection{Results from Simulations}

 We analyse the mock Mg\,II halo {presented in Section~\ref{sec:simu}} in an analogous way to the observations, by calculating the EW fractional difference {of randomly choosen sightline pairs through the halo} for different physical scales. The results are shown in the middle panel of Figure \ref{fig:EWfrac}. We plot sightlines with 0.05 \AA\ $<$ EW $<$ 2 \AA\ in order to match the EW range of the observations. We find that in this EW range, the halo shows indications of both filamentary and clumpy structures in the CGM as well as the outskirts of the galactic disk. We exclude the low density CGM which remains under the typical Mg\,II detection limits and the central part of the galaxy with EW $>$ 2\AA\ which typically consists of several absorbing components in velocity space and are therefore not suitable for comparison with the observations.
 
 Thanks to the high spatial resolution we are confident in the EW fractions at sub-kpc scales and find that we would theoretically expect a similar scatter at those scales as for larger separations.
 However, this is not seen in the observational data and may stem from resolution effects in the observations.
 
In the mock halo, we report a clear trend for high EW Mg\,II systems (blue, corresponding to $\rm log(N_{Mg\,II } [cm^{-2}]) \gtrsim 14 $) showing large variation over all scales ($\sim$ 1-100 kpc) {due to their compactness and small covering fraction.}
{Those high EW systems are mostly found in the inner regions of the halo, close to the galaxy, which explains the decrease of low fractional difference points at large separations.}
The outflowing clumps identified in Figure \ref{fig:foggiehalo} with 13 $ \rm < log(N_{Mg\,II} [cm^{-2}]) <$ 14 have equivalent widths of $\sim$ 0.3 $-$ 0.5 \AA , {which is in the range we observe for the $z_{\rm abs}$=1.393 Mg\,II system}. 
Figure \ref{fig:foggiehalo} shows that the gas with 12 $ \rm < log(N_{Mg\,II} [cm^{-2}]) <$ 13 arises in a mixture of geometries, with filaments dominating over clumps. 
Those filaments have EW values of $<$ 0.2 \AA\ {(similar to our observed system at $z_{\rm abs}$=1.168)} and seem to show the largest coherence lengths of up to $\sim$ 100 kpc.
Low EW Mg\,II systems (green, EW $<$ 0.5 \AA ) show a large scatter in fractional variation at all scales, indicating that those probe a variety of structures, from clouds to filaments with large coherence lengths over tens of kpc.
However, the apparent coherence length of up to 100 kpc from random pairs of sightlines is misleading as those are most likely probing independent, separated structures.

To further investigate the coherence lengths of different CGM structures (clumps, filaments), we compute the fractional EW difference in clumps and filaments separately (right panel of Figure \ref{fig:EWfrac}). 
We limit the data points for this plot to strongly varying systems (EW fractional difference $>$ 0.5) and to spatial scales of up to 6.3 kpc where the scatter in the middle panel of Figure \ref{fig:EWfrac} increases.
Clumps are found to vary strongly almost equally over all those scales, whereas filaments show a trend of stronger variation with larger separation, indicating coherence over larger scales, as illustrated by the column density map in Figure \ref{fig:foggiehalo}. 
This therefore confirms that absorption systems that vary strongly on scales of $<$2 kpc most likely trace clumps, rather than the edges of a filament. {This result allows us to interpret the nature of the Mg\,II systems (see Sections~\ref{sec:nature_z1p167} and 
\ref{sec:nature_z1p392}).}

\cite{Nelson2020} have recently studied the formation and distribution of Mg\,II clouds in Illustris TNG50 for a similar halo in terms of halo mass.
Surprisingly, their Mg\,II absorbers show a homogeneous distribution of small clouds around the halo, which is different from what we observe in the FOGGIE halo.
The two works agree that the majority of Mg\,II systems trace inflows, however there are clear morphological differences in Mg\,II systems between the two simulations. 
This may be due to different physics implemented in the simulations (e.g. magnetic fields {or different feedback models}).
{Indeed, the dependence of the CGM structure on the physics implemented in a simulation has recently been demonstrated by \cite{vandeVoort2020} who show that the distribution of cool gas and mixing of metals is affected by magnetic fields and feedback. 
Future investigations of cool clumps in the CGM, both in simulations and observations are therefore needed to draw further conclusions on the nature of CGM clouds.}

\subsection{Nature of the Observed Systems}

\subsubsection{Lens Galaxy at $z_{\rm lens}$=0.657}

The quadruple quasar images are known to be produced by a foreground massive galaxy \citep{Eigenbrod2006}. Recently, \cite{Sluse2019} have investigated the environment of this lensed quasar and spectroscopically determined the lens redshift to be $z_{\rm lens}$=0.657.
The fiducial lens model for this system with Einstein radius $\theta_E$ = 0.944 arcsec, suggests a halo mass of $M \approx 10^{11.5} M_{\odot}$ \citep{Rusu2019}. 
Recent studies, like the COS-LRG survey \citep{Chen2018,Zahedy2019}, commonly find {cold}, metal enriched gas in the CGM of massive, elliptical and typically quenched galaxies.
{However, in our data }
no tracers of {cold} gas ($\rm log(N_{Mg\,I})<11.1~cm^{-2}$, $\rm log(N_{Ca\,II~H+K})<11.7~cm^{-2}$) are detected against any of the four quasar images. 
This result implies that there is little 10$^4$ K gas around this massive galaxy. 
{We note, however, that the Mg\,I 2852 line is weaker than Mg\,II 2796 which is not covered by the MUSE spectrum and that Ca\,II~H+K usually traces ISM gas. A detection of Ca\,II~H+K would likely indicate a satellite galaxy rather than CGM gas clouds.}
In addition, we do not detect [O\,II] emission associated with the lens, indicating a low SFR, as expected from a quenched elliptical galaxy.

\subsubsection{Mg\,II Absorber at $z_{\rm abs}$=1.168}
\label{sec:nature_z1p167}

The Mg\,II absorber at $z_{\rm abs}$=1.168 is clearly detected against all four quasar images. The absorption feature is particularly strong against counter-image B, with a rest-frame equivalent width EW$_{2796}$=0.15$\pm$0.02\AA. The absorption is weaker towards the other 3 images with EW$_{2796}$ ranging from 0.05$\pm$0.01\AA\ to 0.09$\pm$0.01\AA. Individual values {for this weak (EW(2796)$<$0.3 {\AA}) absorption system} are reported in Table~\ref{tab:onetable}. {The detection of Mg\,II absorption towards all four quasar images implies a cloud coherence length of at least 6 kpc for this absorber. }

Additionally, we detect an [O\,II] emitting galaxy at the redshift of the absorber. 
{The emititer, situated 89 kpc from the absorber, is a low-mass object (stellar mass of M$_\star$ = $10^{9.6\pm0.2}$ M$_{\odot}$) and has a star formation rate (uncorrected for dust-depletion) of SFR $\geq$ 4.6 $\pm$ {1.5} $\rm M_{\odot}$/yr.}
{We find} a fairly high value compared to SFRs of galaxies associated with typical heavy-element absorption systems at these redshifts (e.g. \citealp{kulkarni06,Peroux2011,Augustin2018}). The physical separation indicates that the absorption feature likely probes gas in the CGM of that foreground galaxy. In velocity space, we find the absorber to be blueshifted with respect to the host galaxy {but the measurements are not robust enough to draw further conclusions on the nature of the absorbing cloud}. 

The canonical view is that inflows preferentially occur along the major axis, while outflows are aligned with the minor axis, corresponding to the path of least resistance \citep{Peroux2011, Bouche2012, Schroetter2019, Peroux2020}. The so-called "azimuthal angle" between the quasar line-of-sight and the projected galaxy’s major axis on the sky therefore in principle constrains this alignment. In this case, the azimuthal angle of 45$\pm$41 degrees does not indicate a high likelihood of either accretion or outflow. Given the distance to the [O\,II] emitter and the strength of the Mg\,II absorption, the absorber could also be associated with another closeby dwarf galaxy. However, due to the lack of nearby [O\,II] emission and no clear signs of rotation in the absorption velocities, we do not confirm this hypothesis.

The system at $z_{\rm abs}$=1.168 which consists of weak measured Mg\,II EWs shows little variation in EW fractional difference up to $\sim$ 5-6 kpc. Based on the simulation snapshot, these properties indicate that the absorber is probing a filamentary structure, potentially from gas infalling onto the galaxy. 
Indeed, a qualitative description of the simulations (bottom panel of Figure~\ref{fig:foggiehalo}) indicates that the spatially homogeneous filamentary structures are aligned with inflow dominated regions, while the highly clumpy structures trace outflowing regions. 
The inflowing filaments in Figure \ref{fig:foggiehalo} show Mg\,II column densities of $\rm log(N_{Mg\,II} [cm^{-2}])\sim 12$ with little spatial variation, in line with our observations of the absorber at $z_{\rm abs}$=1.168.
{Given the qualitative consistency of our observed data with simulated inflow regions, we} therefore suggest that the absorber may trace an inflowing filament onto the [O\,II] emitting galaxy.

\subsubsection{Mg\,II Absorber at $z_{\rm abs}$=1.393}
\label{sec:nature_z1p392}
	
The Mg\,II absorber at $z_{\rm abs}$=1.393 is clearly detected against image C, tentatively towards image B but not in the two other quasar image spectra (A1 and A2), with EW$_{2796}$ ranging from 0.05$\pm$0.01\AA\ in image B to 0.41$\pm$0.04\AA\ in image C. 
Individual measurements are reported in Table~\ref{tab:onetable}.  

At the absorber redshift, the physical separations between counter-images B and C is 2.3 kpc and that between A1 and C is 1.7 kpc. (Table~\ref{tab:EWfracs}). At the redshift of the Mg\,II absorbers, no galaxies are detected down to {SFR < 0.02 M$_{\odot}$/yr}, 
so that the location of the gas with respect to a bound system is not known. Given than the Mg\,II absorption line is solely detected against quasar images C and B, we assume that the metal cloud is centered close to image C. This configuration puts strong constraints on the {coherence length of the cold} gas cloud associated with this Mg\,II absorber. Given the non-detections against counter-images A1 and A2, we conclude that the density of the Mg\,II cloud varies significantly over scales smaller than 2 kpc (between C and A1). 

{The system at $z_{\rm abs}$=1.393 is a large equivalent width absorber against one quasar image but marginally or no detected towards the others. } The system thus presents sizeable EW fractional differences. {With the additional insights from simulations, we infer that the absorber} is likely probing one of the numerous small, dense {outflowing} CGM gas clumps (Figure \ref{fig:foggiehalo}).

	\section{Conclusions}

Here, we present new 3D spectroscopy VLT/MUSE observations of a gravitationally lensed quasar displaying four bright images. The adaptive optics system of MUSE provides the necessary spatial resolution to enable the extraction of individual spectra of each of the quasar images. To reach deep observations without losing the spatial resolution, we have carefully combined our AO observations with archival observations by weighting individual exposures on their image quality. {The combination of the cubes is performed by assigning  a larger weight to longer exposure times and better seeing values.} In this work, we systematically search for foreground absorption features, both at the known redshift of the lens and along the entire line-of-sight. 
We detect [O\,II] emission associated with one of the Mg\,II absorbers. We further present specifically designed cosmological hydrodynamical zoom-in FOGGIE simulations of an equivalent Mg\,II halo with enhanced spatial resolution in the CGM in order to capture the small-scale physics and substructure of the bulk of the gas in the galactic ecosystem. The combination of observations and simulations put new constraints on the spatial {clumpiness of cold} (T = 10$^4$K) gas around galaxies. 

Our main results {can be summarised as follows}:

	\begin{itemize}

	\item {We report the non detection of cold} 10$^4$ K gas around the massive ($M \approx 10^{11.5} M_{\odot}$) $z_{\rm lens}$=0.657 lens galaxy. 

	\item We {calculate} the EW fractional difference {of both the new observations together with an extended literature sample and the simulated Mg\,II halo. We find that fractional difference} depends on the underlying structure of the absorber, with extended inflow filaments showing coherence {length} on larger scales than outflowing clumpy structures. 

	\item {We show that} the $z_{\rm abs}$=1.168 Mg\,II system possibly probes an inflowing gas filament onto an [O\,II] emitting galaxy situated 89 kpc away with SFR $\geq$ 4.6 $\pm$ {1.5} $\rm M_{\odot}$/yr and $M_{*}=10^{9.6\pm0.2} M_{\odot}$. 

	\item {Using similar lines of argument, we establish that the physical properties of the $z_{\rm abs}$=1.393 Mg\,II system indicate that this absorber} likely traces dense CGM gas clumps varying in strength over $\sim$ 2 kpc physical scales.

	\end{itemize}

Enlarged samples of multiply lensed bright quasars observed with IFU instruments combined with state-of-the-art zoom-in hydrodynamical simulations are essential to put new constraints on the {clumpiness} of the gas in the CGM of galaxies, a key diagnostic of the baryon cycle.

\section*{Acknowledgements}

We thank Kate Rubin for sharing data and Molly Peeples, Jason Tumlinson and the {\it FOGGIE\footnote{https://foggie.science}} collaboration for making their simulations available and for related discussions. This work has made use of the yt\footnote{https://yt-project.org} package \citep{Turk2011}.
RA was supported by NASA grants 80NSSC18K1105 and HST GO \# 15075.
VPK acknowledges partial support from NASA grants NNX17AJ26G and 80NSSC20K0887.
This research made use of Astropy,\footnote{http://www.astropy.org} a community-developed core Python package for Astronomy \citep{astropy:2013, astropy:2018}.

\section*{Data Availability}
All data incorporated into the article are available on request.



\bibliographystyle{mnras}
\bibliography{biblio} 

\bsp	
\label{lastpage}
\end{document}